\documentclass[aps,pra,twocolumn,showpacs,superscriptaddress]{revtex4}
\usepackage{graphicx}
\usepackage{epstopdf}
\usepackage{amsmath}
\usepackage{amssymb}
\usepackage{float}
\usepackage{bm}
\usepackage{threeparttable}

\renewcommand{\u}[1]{\underline{#1}}

\input{epsf}
\begin{document}

\title{Scattering properties of the $2e^-2e^+$ polyelectronic system}

\author{K.~M. Daily}
\affiliation{Department of Physics,
Purdue University,
West Lafayette, Indiana 47907, USA}
\author{Javier von Stecher}
\affiliation{Seagate Technology,
Longmont, Colorado 80503, USA}
\author{Chris H. Greene}
\affiliation{Department of Physics,
Purdue University,
West Lafayette, Indiana 47907, USA}

\date{\today}

\begin{abstract}
We study the $2e^-2e^+$ equal-mass charge-neutral four-body system 
in the adiabatic hyperspherical framework.
The lowest few adiabatic potentials are calculated for 
zero orbital angular momentum, positive parity, 
and charge conjugation symmetries.
Propagating the $R$-matrix,
the low-energy $s$-wave scattering lengths of the singlet-singlet
and triplet-triplet spin configurations are calculated.
Lastly,
we calculate the $S$-matrix for energies above the ionic threshold 
to estimate the transition rates between
the single ionic fragmentation channel 
and the lowest few dimer-dimer fragmentation channels.
\end{abstract}


\maketitle

\section{Introduction}
Many charged few-body systems have been studied using
the hyperspherical framework.
From Macek's pioneering work~\cite{macek1968},
the three-body Coulomb problem has been extensively
studied~\cite{lin1974,botero1986,macek1986,gusev1990,
sadeghpour1990,archer1990,esry2003}
and reviewed (see Refs.~\cite{gusev1990,lin1995} and references within).
There have been notable studies of four-body
charged systems as well~\cite{watanabe1982,morishita1998,dincao2003},
and even a first look at the equal-mass five-body system~\cite{daily2014}.

An interesting class of charged systems
is that of particle-antiparticle mixtures~\cite{hagiwara2002}.
These systems are inherently unstable due to annihilation,
but are fundamental, e.g.,
the high rates of Ps annihilation 
at the galactic bulge~\cite{weidenspointner2006}
or precision tests of CPT violation 
via antihydrogen~\cite{hagiwara2002,andresen2006}.
Moreover,
there have been many advancements towards a gamma-ray laser
based on the development of a BEC of spin-polarized 
Ps~\cite{platzman1994,mills2002,cassidy2005,avetissian2014,wang2014}
and the experimental observation of the Ps$_2$ molecule~\cite{cassidy2007}.

Using the adiabatic hyperspherical framework~\cite{rittenhouse2011,daily2014a},
we examine the $2e^-2e^+$ equal-mass charge-neutral four-body system.
The adiabatic Hamiltonian is solved
utilizing a basis of explicitly correlated Gaussians at a fixed
hyperradius\cite{vonstecher2009,rakshit2012,2013RMPmitroy}.
The evaluation of matrix elements is facilitated using
a Fourier transform technique~\cite{daily2014}.
The hyperradial $R$-matrix is propagated and
the low-energy $s$-wave scattering lengths are calculated.
Also,
from the $S$-matrix we estimate the transition probabilities between the
ionic channel and the lowest few Ps$_2$ dimer-dimer channels.

The rest of the paper is organized as follows.
Section~\ref{sec_theory} defines the Hamiltonian 
and the basis set used to solve the adiabatic Hamiltonian.
Section~\ref{sec_pcurves}
analyzes the adiabatic potential curves,
that is,
the solutions to the adiabatic Hamiltonian.
Section~\ref{sec_scattering} describes how the
low-energy elastic scattering lengths are calculated
and compares the results to benchmark calculations from
the literature.
In Sec.~\ref{sec_charge},
the $S$-matrix is calculated to
estimate the transition probabilities between the lowest
few dimer-dimer fragmentation channels and
the single ionic fragmentation channel.
Last, Sec.~\ref{sec_conclusion} concludes.

\section{Theoretical background}
\label{sec_theory}
Consider the system of two electrons and two positrons
in three dimensions interacting via the two-body Coulomb potential.
The Hamiltonian $H$ in atomic units $(\hbar=m_e=1)$ reads
\begin{align}
H = & 
-\frac{1}
      {2} 
\sum_{j=1}^4 
\nabla^2_{\bm{r}_j}
+\sum_{i<j}
\frac{q_i q_j}
     {|\bm{r}_i-\bm{r}_j|}
\end{align} 
where $\bm{r}_j$ is the location of particle $j$. 
For concreteness,
$q_1=q_2=+1$ and $q_3=q_4=-1$.
The center of mass $H_{\rm CM}$ and relative $H_{\rm rel}$ 
contributions separate,
$H = H_{\rm CM} + H_{\rm rel}$.
Our focus is on the relative Hamiltonian,
\begin{align}
H_{\rm rel}  = 
  - \frac{1}{2 \mu} \sum_{j=1}^3 \nabla^2_{\bm{\rho}_j}
  + V_C(\bm{\rho}_1,\bm{\rho}_2,\bm{\rho}_3),
\end{align}
where $V_C$ contains the pair-wise Coulomb interactions as a function of
the three relative Jacobi vectors $\bm{\rho}_j$,
$j=1,2,3$.
All Jacobi vectors are scaled such that they are analogous to 
three equal-mass ``particles'' of mass $\mu$.
We take $\mu=2^{-2/3}$ such that the coordinate transformation is unitary.

The relative Hamiltonian $H_{\rm rel}$ is recast in hyperspherical
coordinates in terms of eight hyperangles denoted by $\bm{\Omega}$
and a single length, the hyperradius $R$.
The relative Hamiltonian is then a sum of the 
hyperradial kinetic energy ${\cal T}_{R}$,
the hyperangular kinetic energy ${\cal T}_{\bm{\Omega}}$, and
the interaction potential,
\begin{align}
\label{eq_SErel}
H_{\rm rel} = {\cal T}_R + {\cal T}_{\bm{\Omega}} + V_{\rm int}(R,\bm{\Omega}),
\end{align}
where
\begin{align}
{\cal T}_R = - \frac{1}{2 \mu} 
\frac{1}{R^{8}}\frac{\partial}{\partial R}R^{8} 
\frac{\partial}{\partial R}.
\end{align}
The exact form of the hyperangular kinetic energy ${\cal T}_{\bm{\Omega}}$
depends on the choices of the Jacobi vectors and of the hyperangles.
The exact form is not needed here,
but additional detail can be found in Ref~\cite{daily2014a}.

The solution $\Psi_E(R,\bm{\Omega})$ 
to Eq.~\eqref{eq_SErel} is expanded in terms of the
radial functions $R^{-4} F_{E\nu}(R)$ and
the channel functions $\Phi_{\nu}(R;\bm{\Omega})$,
\begin{align}
\label{eq_psi_expansion}
\Psi_E(R,\bm{\Omega}) = 
R^{-4}\sum_{\nu} F_{E\nu}(R)\Phi_{\nu}(R;\bm{\Omega}).
\end{align}
The channel functions at a fixed hyperradius $R$ form
a complete orthonormal set over the hyperangles,
\begin{align}
  \int \,d\bm{\Omega} \; \Phi^*_{\nu}(R;\bm{\Omega}) \Phi_{\nu'}(R;\bm{\Omega}) 
  = \delta_{\nu \nu'},
\end{align}
and are the solutions to the adiabatic Hamiltonian $H_{\rm ad}(R,\bm{\Omega})$,
\begin{align}
  H_{\rm ad}(R,\bm{\Omega}) \Phi_{\nu}(R;\bm{\Omega}) 
  = U_{\nu}(R) \Phi_{\nu}(R;\bm{\Omega}),
\end{align}
where
\begin{align}
\label{eq_Had}
H_{\rm ad}  = & 
\frac{\bm{\Lambda}^2 + 12 }{2\mu R^2} 
+ \frac{C(\bm{\Omega})}{R}.
\end{align}
Here, $\bm{\Lambda}^2$ is the square of the grand angular momentum operator
and $C(\bm{\Omega})$ is the hyperangular part of the Coulomb interaction.

After applying Eq.~\eqref{eq_SErel} on the expansion 
Eq.~\eqref{eq_psi_expansion}
and projecting from the left onto the channel functions,
the Schr\"odinger equation reads
\begin{align}
\label{eq_SE_W}
&
\left( 
  -\frac{1}{2\mu}\frac{d^2}{d R^2} 
  + U_{\nu}(R)  -E
\right) F_{E \nu}(R)
\\ \nonumber & \qquad 
-\frac{1}{2\mu} \sum_{\nu'}
  \left( 2 P_{\nu \nu'}(R)\frac{d}{d R} + Q_{\nu \nu'}(R) \right)
  F_{E \nu'}(R)
= 0.
\end{align}
The hyperspherical Schr\"odinger equation Eq.~\eqref{eq_SE_W}
is solved in a two step procedure.
First, $H_{\rm ad}(R,\bm{\Omega})$ is solved parametrically in $R$ for
the adiabatic potential curves $U_{\nu}(R)$.
In a second step,
the coupled set of one-dimensional equations in $R$ are solved. 
In Eq.~\eqref{eq_SE_W}, $P_{\nu\nu'}$ and $Q_{\nu\nu'}$ 
represent the coupling between channels,
where 
\begin{align}
\label{eq_Pcoupling}
  P_{\nu \nu'}(R) = \bigg\langle \Phi_{\nu} \bigg| 
  \frac{\partial \Phi_{\nu'}}{\partial R} \bigg\rangle_{\bm{\Omega}}
\end{align}
and
\begin{align}
\label{eq_Qcoupling}
  Q_{\nu \nu'}(R) = \bigg\langle \Phi_{\nu} \bigg| 
  \frac{\partial^2 \Phi_{\nu'}}{\partial R^2} \bigg\rangle_{\bm{\Omega}}.
\end{align}
The brackets indicate that the integrals are taken only over the
hyperangle $\bm{\Omega}$ with the hyperradius $R$ held fixed.

The eigenfunctions $\Phi_{\nu}(R;\bm{\Omega})$ 
of $H_{\rm ad}(R,\bm{\Omega})$ are simultaneous eigenstates
of the total orbital angular momentum $L$,
the parity $\pi$,
and the spin of the identical positrons $S_+$
and identical electrons $S_-$.
Moreover,
because of the equal masses and charges,
the eigenfunctions are also eigenstates of the charge conjugation
operator $\hat{C}$.
The $\Phi_{\nu}(R;\bm{\Omega})$ 
are expanded using
a non-orthogonal basis of correlated Gaussians~\cite{vonstecher2009,rakshit2012,2013RMPmitroy},
\begin{align}
\label{eq_basis}
|\Phi_{\nu}\rangle = \sum_j \hat{C}_{\pm}\hat{S} |A^{(j)} \rangle |\chi\rangle,
\end{align}
where $\hat{S}$ is a symmetrization operator
that permutes the space and spin labels of identical particles.
In particular,
$\hat{S}=[1-(12)][1-(34)]$,
where $(ij)$ is the two-cycle operator that exchanges particles $i$ and $j$.
The three-cycle operator $(ijk)$,
for example,
denotes the permutation $i \to j$, $j \to k$, and $k \to i$.
In practice,
to project out the parts of the functions 
that are either even ($+$) or odd ($-$) under charge conjugation,
we apply the operator
$\hat{C}_{\pm}$, $\hat{C}_{\pm}=1 \pm \hat{C}$, 
where $\hat{C}=(13)(24)$.

In general,
under permutation
the spin functions $|\chi \rangle$
would transform to a different spin configuration.
However,
this work only considers the singlet-singlet (SS)
or triplet-triplet (TT) spin configurations.
In this case,
the effect of permutations on the spin functions 
leaves them unchanged except possibly for an overall minus sign.
The combined operator $\hat{C}_{\pm}\hat{S}$
involves eight permutations.
Table~\ref{tab_perm}
\begin{table}
\caption{Permutations used in the basis functions Eq.~\eqref{eq_basis}
labeled by the charge conjugation and spin.
SS (TT) means singlet-singlet (triplet-triplet).
}
\begin{ruledtabular}
\begin{tabular}{lllll}
         & $C_+$SS & $C_+$TT & $C_-$SS & $C_-$TT \\
1        & $+$ & $+$ & $+$ & $+$ \\
(12)     & $+$ & $-$ & $+$ & $-$ \\
(34)     & $+$ & $-$ & $+$ & $-$ \\
(12)(34) & $+$ & $+$ & $+$ & $+$ \\
(13)(24) & $+$ & $+$ & $-$ & $-$ \\
(3142)   & $+$ & $-$ & $-$ & $+$ \\
(1324)   & $+$ & $-$ & $-$ & $+$ \\
(14)(23) & $+$ & $+$ & $-$ & $-$
\end{tabular}
\end{ruledtabular}
\label{tab_perm}
\end{table}
indicates all of the permutations 
and their effective signs that are considered in this work.
The first column denotes the permutation
while the first row denotes the system considered.

The functions $|A^{(j)} \rangle$ are
\begin{align}
\label{eq_sph_gauss}
|A^{(j)}\rangle = \exp\left( -\frac{1}{2} \bm{x}^T\u{A}^{(j)}\bm{x} \right)
\left| \bm{u}^T \bm{x} \right|^{2K}. 
\end{align}
Here, $\bm{x}$ is an array of (column) Jacobi vectors, 
$\bm{x}^T=\{\bm{x}_1,\bm{x}_2,\ldots,\bm{x}_N  \}$.
All Jacobi vectors exist in three dimensions,
such that the $j^{th}$ Jacobi vector reads
$\bm{x}_j^T=\{ x_{j,1},x_{j,2},x_{j,3} \}$.
$\u{A}^{(j)}$ is an $N \times N$ symmetric positive definite 
coefficient matrix that describe the correlations.
The matrix $\u{A}^{(j)}$ contains $N(N+1)/2$ 
independent variational parameters.
The $N$-dimensional global vector $\bm{u}$
determines the linear combination of Jacobi vectors,
where the integer $K$ is a nodal parameter. 
These basis functions describe only natural parity
states $[\pi=(-1)^L]$ with zero orbital angular momentum $L$, 
though it is well-known how to extend this basis to include
unnatural parity~\cite{2013RMPmitroy}.

\section{Adiabatic potential curves}
\label{sec_pcurves}

\begin{figure}
\vspace*{+1.5cm}
\centering
\includegraphics[angle=0,width=70mm]{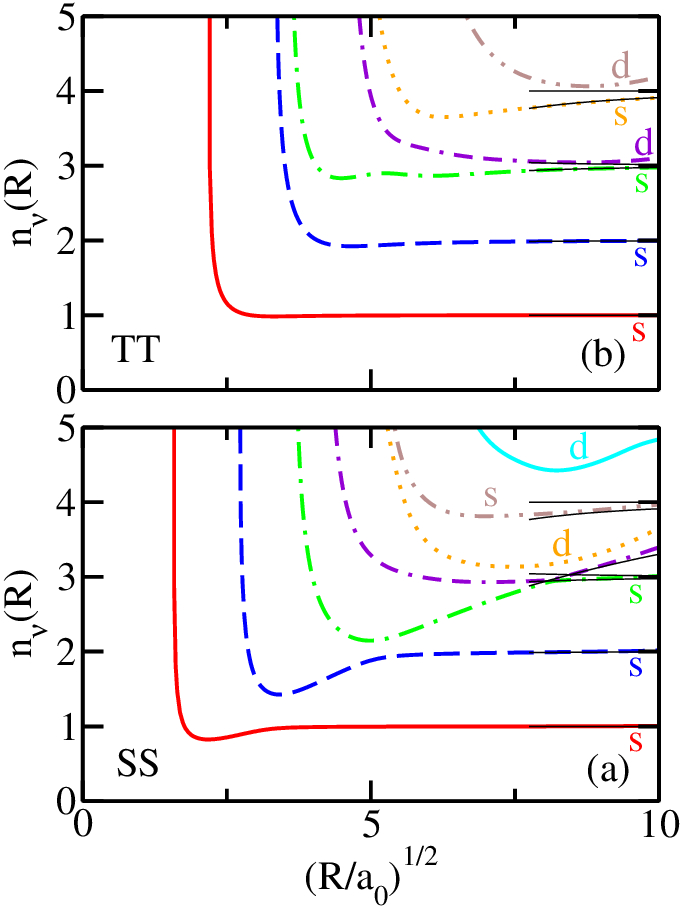}
\vspace*{0.1cm}
\caption{(Color online) 
Adiabatic potential curves for $L^{\pi}=0^+$ 
and charge conjugation eigenvalue $+1$ 
shown as effective quantum numbers [see Eq.~\eqref{eq_effqn}]
vs $\sqrt{R}$.
Panels (a) and (b) are for $(S_+,S_-)=(0,0)$ and
$(1,1)$, respectively.
The thin solid lines show the known asymptotic behavior 
through order $R^{-3}$.
The asymptotically ionic channel in (a) is the dash-dash-dotted line.
The dimer-dimer asymptotic thresholds are labeled by
the angular momentum of the excited Ps.
}\label{fig_CplusSSTT_scaled}
\end{figure}

\begin{figure}
\vspace*{+1.5cm}
\centering
\includegraphics[angle=0,width=70mm]{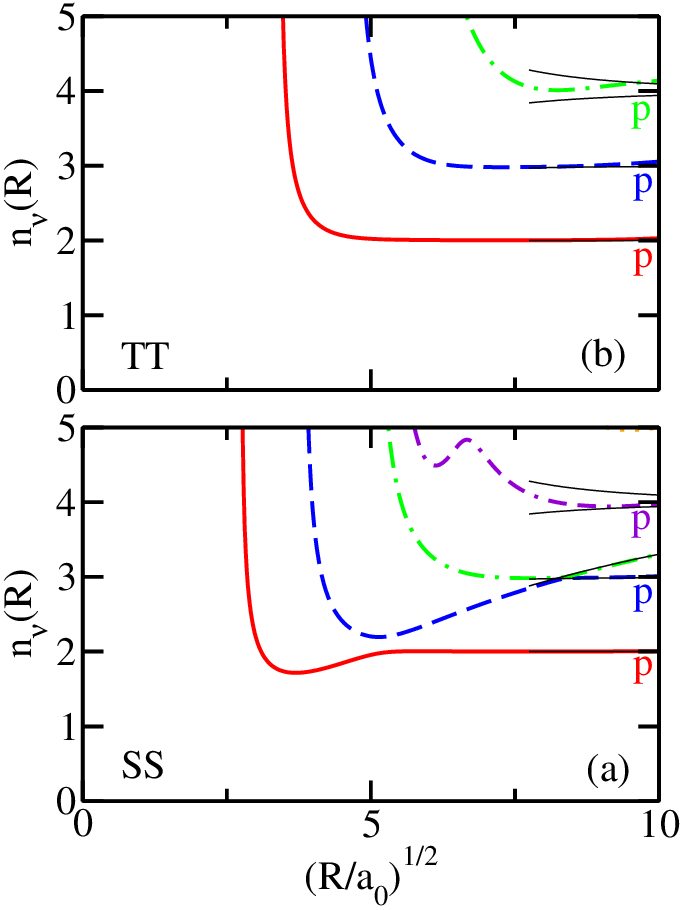}
\vspace*{0.1cm}
\caption{(Color online) 
Adiabatic potential curves for $L^{\pi}=0^+$ 
and charge conjugation eigenvalue $-1$ 
shown as effective quantum numbers vs $\sqrt{R}$.
Panels (a) and (b) are for $(S_+,S_-)=(0,0)$ and
$(1,1)$, respectively.
The thin solid lines show the known asymptotic behavior 
through order $R^{-3}$.
The asymptotically ionic channel in (a) is the dash-dotted line.
The dimer-dimer asymptotic thresholds are labeled by
the angular momentum of the excited Ps.
}\label{fig_Cminus_scaled}
\end{figure}

This paper is concerned with states of $L=0$ 
angular momentum and positive parity $\pi$.
Our matrix element calculations
utilize a technique that reduces all matrix element evaluations 
to 1-D Fourier transforms~\cite{daily2014}.
Using this basis,
the lowest few adiabatic potentials are calculated for
the $2e^-2e^+$ system and shown in Figs.~\ref{fig_CplusSSTT_scaled}
and~\ref{fig_Cminus_scaled}.
To put all adiabatic potentials on the same scale,
the curves are plotted as effective quantum numbers 
$n_{\nu}(R)$~\cite{sadeghpour1990,esry2003}
as functions of the square root of the hyperradius $R$.
This scaling is motivated by the fact that, at low energy,
all curves except the asymptotically ionic channel break up
into two Ps dimers at large hyperradius.
At the low energies considered here,
one of these Ps dimers is always in the ground state
in the asymptotic limit.
Thus,
based on the asymptotic thresholds energies,
we define
\begin{align}
\label{eq_effqn}
n_{\nu}(R) = \left[-4U_{\nu}(R)/E_H-1 \right]^{-1/2}
\end{align}

Figure~\ref{fig_CplusSSTT_scaled} shows the eigenstates of $\hat{C}$
corresponding to eigenvalue $+1$
for (a) the SS  
and (b) the TT  
spin configurations.
In general,
the potentials of the TT  
case are more repulsive
since the Pauli exclusion principle keeps the identical particles
further apart.
Moreover,
only in the SS  
case does the ionic channel appear.
In a diabatic picture,
this ionic channel, with threshold energy $-0.262E_H$ 
[$n_{\nu}(\infty) = 4.56$],
crosses dimer-dimer channels with $n_{\nu}(\infty) \le 4$,
each crossing becoming less sharp (less diabatic) 
for crossings at lower energy.

Figure~\ref{fig_Cminus_scaled} shows the eigenstates of $\hat{C}$
corresponding to eigenvalue $-1$
for (a) the SS  
and (b) the TT  
spin configurations.
Again,
we observe the potentials of the TT case are more repulsive
that the potentials of the SS spin configuration.
Panel (a) shows the diabatic-like ionic channel,
which is absent in panel (b).
Moreover,
the dash-dash-dotted line of panel (a)
has a local minimum around $\sqrt{R/a_0}\approx 6.5$.
Though not visible on the scale shown,
this is due to an avoided crossing with the next highest channel.
Though the curves begin to be less converged in this region,
it is a true feature and not an artifact.
Taking into account the other adiabatic potentials (not shown),
even though unconverged,
they hint that this is just the first of many avoided crossings
at large $R$ and appear to map out a diabatic curve
that asymptotically approaches $-0.25E_H$,
that is,
the energy where one dimer has completely dissociated.

The asymptotic limits up through order $R^{-3}$ are shown as
thin solid lines at large $R$ 
in both Figs.~\ref{fig_CplusSSTT_scaled} and~\ref{fig_Cminus_scaled}.
This asymptotic behavior can be calculated by asymptotically
expanding the adiabatic Hamiltonian
in powers of $R^{-1}$~\cite{daily2014a}
and using degenerate perturbation theory.
This yields, 
for the dimer-dimer channels shown,
\begin{align}
& U_{\nu}(R \to \infty) \approx  
\frac{-1}{4} + \frac{-1}{4n^2}
\nonumber \\ &
+\frac{1}{4\mu R^2} \bigg[3l(l+1)-n^2-2
                          - \frac{2^8 n^5 \delta_{l0}}{(n^2-1)^4} 
                          \left(\frac{n-1}{n+1}\right)^{2n}\bigg]
\nonumber \\ &
+\frac{1}{3\mu^{3/2}R^3} \frac{2^{11} n^7 \delta_{l1}}{(n^2-1)^5}
    \left(\frac{n-1}{n+1}\right)^{2n},
\end{align}
where the two terms involving Kronecker deltas
only contribute if $n>1$.

The number of asymptotic channels can be understood
by examining the asymptotic wave functions~\cite{daily2014a}.
For the dimer-dimer thresholds,
ignoring the spin part of the wave function,
the unsymmetrized asymptotic wave function is effectively a product of
two scaled hydrogenic radial wave functions and a coupled set
of spherical harmonics whose angles are defined 
by the Jacobi vectors $\bm{\rho}_j$,
\begin{align}
|\Phi(R\to\infty)\rangle \approx |n_1l_1\rangle_1 |n_2l_2 \rangle_2 
|\hat{\rho}_1\hat{\rho}_2\hat{\rho}_3\rangle,
\end{align} 
where $|nl\rangle_j$ represent the hydrogenic wave function along 
the $jth$ Jacobi vector and
\begin{align}
|\hat{\rho}_1\hat{\rho}_2\hat{\rho}_3\rangle = &
\frac{(-1)^{l_3}}{\sqrt{2l_3+1}} \sum_{\bm{m}}
\langle l_1 m_1l_2m_2|l_3m_3 \rangle \times
\nonumber \\
& Y_{l_1m_1}(\hat{\rho}_1)Y_{l_2m_2}(\hat{\rho}_2)Y^*_{l_3m_3}(\hat{\rho}_3).
\end{align}
The sum is over all projection quantum numbers,
$Y$ are spherical harmonics,
and $\langle \cdot \rangle$ is a Clebsch-Gordan coefficient.

The Jacobi vector $\bm{\rho}_1$ defines the first dimer,
$\bm{\rho}_2$ defines the second dimer,
and $\bm{\rho}_3$ defines the inter-dimer distance.
Applying the symmetrization operator $\hat{S}$ yields
\begin{align}
\label{eq_phisymmetrized}
\hat{S}|\Phi(R\to\infty)\rangle \approx &
|n_1l_1\rangle_1 |n_2l_2 \rangle_2 |\hat{\rho}_1\hat{\rho}_2\hat{\rho}_3\rangle
\nonumber \\ &
+(-1)^{l_3}
|n_2l_2\rangle_1 |n_1l_1 \rangle_2 |\hat{\rho}_2\hat{\rho}_1\hat{\rho}_3\rangle
\end{align}
since the terms arising from $P_{12}$ and $P_{34}$ are exponentially negligible.
This can be understood since,
if expressed in a single Jacobi basis,
these operators cause the Jacobi vectors to
pick up components along $\bm{\rho}_3$.
In the asymptotic limit,
$\bm{\rho}_3$ scales with the hyperradius $R$
and thus the hydrogenic wave function causes the exchange term
to vanish exponentially.
The functions Eq.~\eqref{eq_phisymmetrized} are eigenstates of
the charge conjugation projection operator $\hat{C}_{\pm}$,
\begin{align}
\hat{C}_{\pm} \hat{S} |\Phi(R\to\infty)\rangle = 
\left[1 \pm (-1)^{l_1+l_2}\right]\hat{S} |\Phi(R\to\infty)\rangle.
\end{align}
If one dimer is in the ground state,
then the asymptotic wave function vanishes for $\hat{C}_+$
if the other dimer is in an odd partial wave.
On the other hand,
the asymptotic wave function vanishes for $\hat{C}_-$
if the other dimer is in an even partial wave.

In this way,
in a diabatic picture
the adiabatic potentials can be labeled
by how they approach the asymptotic thresholds
(shown as thin solid lines at large $R$).
In Fig.~\ref{fig_CplusSSTT_scaled}(b),
for example,
solid, dashed, dash-dotted, dash-dash-dotted, dotted, and dash-dot-dotted lines
are for $1s1s$, $1s2s$, $1s3s$, $1s3d$, $1s4s$, and $1s4d$ Ps-Ps channels, 
respectively.
In Fig.~\ref{fig_CplusSSTT_scaled}(a),
the labeling is shifted due to the inclusion of 
the ionic channel, 
such that
the ionic channel is the dash-dash-dotted line while
solid, dashed, dash-dotted, dotted, dash-dot-dotted, and solid lines
are for $1s1s$, $1s2s$, $1s3s$, $1s3d$, $1s4s$, and $1s4d$ Ps-Ps channels, 
respectively.
In Fig.~\ref{fig_Cminus_scaled}(b),
solid, dashed, and dash-dotted lines
are for $1s2p$, $1s3p$, and $1s4p$ Ps-Ps channels, 
respectively.
In Fig.~\ref{fig_Cminus_scaled}(a),
the labeling is shifted due to the inclusion of 
the ionic channel,
such that
the ionic channel is the dash-dotted line while
solid, dashed, and dash-dash-dotted lines
are for $1s2p$, $1s3p$, and $1s4p$ Ps-Ps channels, 
respectively.

Our current scheme suffers from convergence issues
in the asymptotically large $R$ region
for states other than $s$-wave.
This is visible in the dotted and upper solid lines 
of Fig.~\ref{fig_CplusSSTT_scaled}(a),
the dash-dash-dotted and dash-dot-dotted lines 
of Fig.~\ref{fig_CplusSSTT_scaled}(b),
as well as the dash-dotted line 
of Fig.~\ref{fig_Cminus_scaled}(b).
Even more,
the potentials that asymptotically approach the $1s4f$ threshold
in Fig.~\ref{fig_Cminus_scaled} are even less converged,
not appearing within the figure on the scale shown.
In practice,
the adiabatic potentials are smoothly matched to the
known asymptotic behavior.

\section{Low energy elastic scattering}
\label{sec_scattering}
This section describes the low energy elastic scattering properties
for those systems shown in Fig.~\ref{fig_CplusSSTT_scaled},
that is,
the $s$-wave scattering lengths for the TT 
and SS  
spin configurations.
The inverse log-derivative $R$-matrix is propagated 
from small hyperradius out to some matching distance $R_m$,
where it is matched to the asymptotic form of the hyperspherical
wave functions.
For short-range interaction potentials,
the couplings and adiabatic potentials fall off sufficiently fast
such that a sufficiently large matching point $R_m$
leads to converged results.
For the long-range Coulomb interaction, however,
we find it better to match the $R$-matrix at many different points
and then extrapolate to infinite matching point.

\begin{figure}
\vspace*{+1.5cm}
\centering
\includegraphics[angle=0,width=70mm]{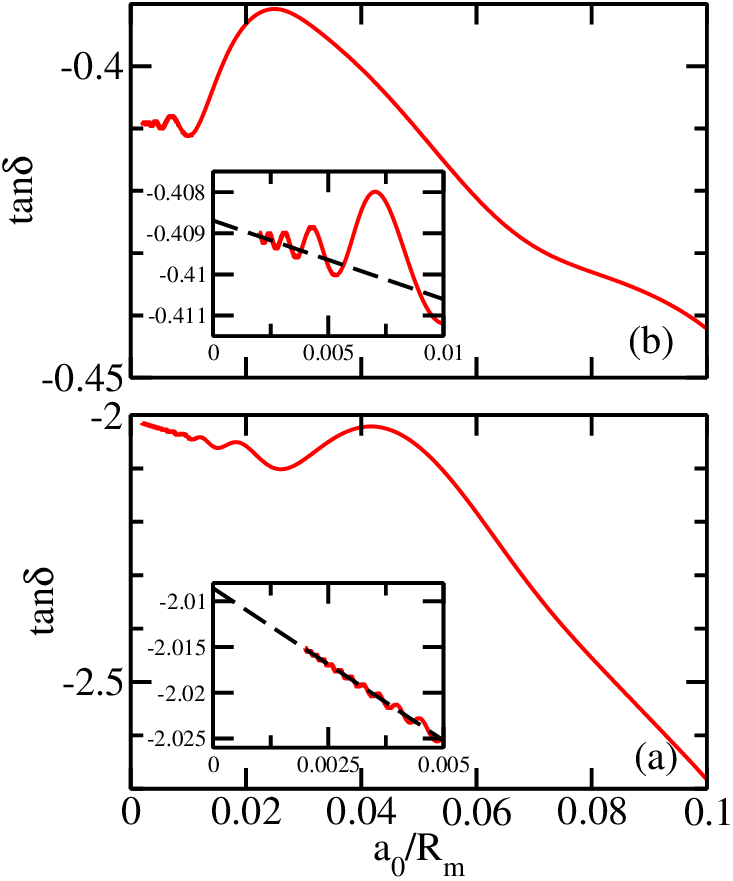}
\vspace*{0.1cm}
\caption{(Color online)
Tangent of the phase shift $\delta$ 
as a function of the matching distance $R_m$
for low-energy elastic scattering of the
$(S_+,S_-)=(0,0)$, $L^{\pi}=0^+$ system using the four lowest channels.
Panels (a) and (b) are for a scattering energies 
of $E_{\rm scatt} = 0.01E_H$ and
$0.001E_H$, respectively.
The dashed line of the insets shows a linear fit over $a_0/R_m=0-0.003$. 
}\label{fig_phaseshift}
\end{figure}

As an example,
Fig.~\ref{fig_phaseshift}
shows the tangent of the $s$-wave phase shift
as a function of the inverse of the matching point
for the SS  
system.
The TT  
system is qualitatively similar.
Panel (a) is for a scattering energy $E_{\rm scatt}=0.01E_H$ above the $1s1s$
threshold,
while panel (b) is for a scattering energy $E_{\rm scatt}=0.001E_H$.
The oscillations in $\tan\delta$ as a function of $R_m$
begin at $R\approx 25a_0$,
that is,
beyond the distance where there is an appreciable potential well.
However,
it is the long-range nonadiabatic coupling,
which between $s$-wave dimer-dimer states 
goes as $P_{\nu\nu'}(R\to\infty)\approx R^{-1}$ at large distance,
that leads to the oscillating behavior.
This has been verified by artificially turning off these couplings
beyond some large distance 
and observing that the oscillations cease.

In addition to the oscillating behavior,
$\tan\delta$ approaches the infinite matching point linearly
when plotted as a function of $R_m^{-1}$.
In practice,
we fit to this linear behavior,
making the range of the fit extend over many wavelengths
to average out the oscillations.
The insets of Fig.~\ref{fig_phaseshift} show such fits.
The oscillations have a smaller wavelength as the scattering energy increases,
such that the fits to the $\tan\delta$ with the lowest scattering energy
have a larger uncertainty.

\begin{figure}
\vspace*{+1.5cm}
\centering
\includegraphics[angle=0,width=70mm]{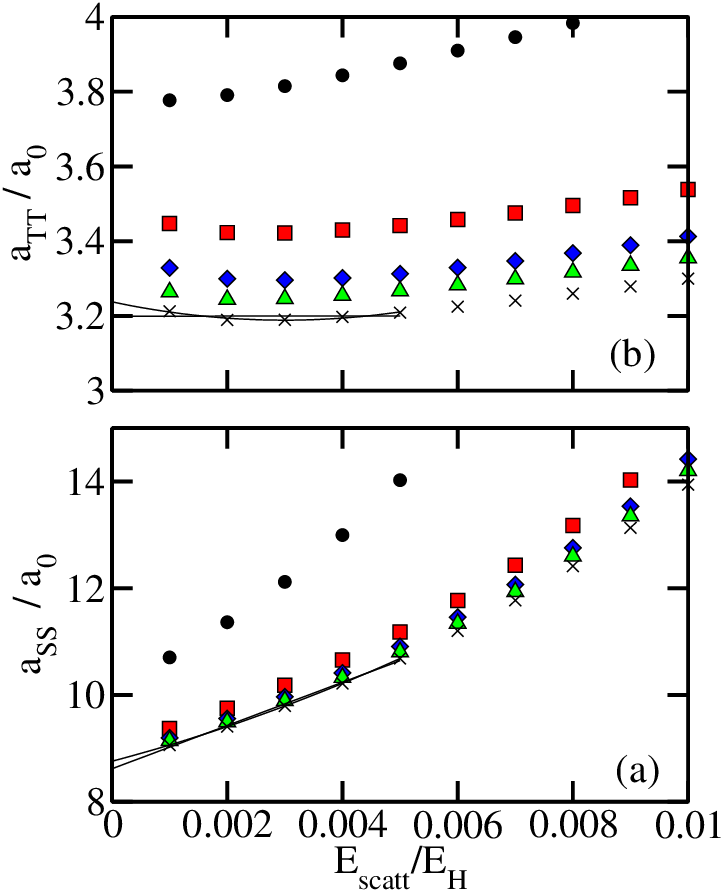}
\vspace*{0.1cm}
\caption{(Color online)
Low-energy $s$-wave scattering length $a$ 
vs scattering energy $E_{\rm scatt}$ for
(a) the SS  
configuration and
(b) the TT  
configuration.
Circles, squares, diamonds, and triangles are for $1-4$ included channels,
respectively.
Crosses represent extrapolating to infinite number of channels,
based on the data for $2-4$ included channels. 
Upper and lower thin solid lines are quadratic and linear fits,
respectively, of the extrapolated data.
}\label{fig_elasticscattering}
\end{figure}

Figure~\ref{fig_elasticscattering}
shows the energy-dependent $s$-wave scattering lengths, 
$a = -\tan\delta/k$,
where $k$ is the scattering momentum,
as a function of the scattering energy $E_{\rm scatt}$,
where $E_{\rm scatt}$ is the energy above the lowest dimer-dimer
threshold of $-0.5E_H$.
Circles, squares, diamonds, and triangles show the
show the SS 
$s$-wave scattering length $a_{\rm SS}$ including
one, two, three, or four lowest channels.
The crosses are an extrapolation to an infinite number of channels
based on the data for including the lowest two through four channels.
The thin lines show linear and quadratic fits to the extrapolated
data set using the five points of lowest scattering energy.
In Fig.~\ref{fig_elasticscattering}(b),
a similar analysis is done 
for the TT  
scattering length $a_{\rm TT}$.
The analysis estimates the scattering lengths to be
$a_{\rm SS} = 8.7(2)a_0$ and $a_{\rm TT}=3.2(1)a_0$.

The scattering estimates provided agree well with others
(see Table~II), 
but are systematically large
when compared 
to the accurate stochastic variational method~\cite{ivanov2001,ivanov2002}.
\begin{table}
\label{tab_scatt}
\begin{threeparttable}[b]
\caption{The $s$-wave scattering length in atomic units for some
calculations of Ps-Ps scattering.}
\begin{ruledtabular}
\begin{tabular}{lcc}
Method & Singlet & Triplet \\ \hline
CCA~\cite{adhikari2002}\tnote{a} & 7.46 & 1.56 \\
CCA~\cite{chakraborty2004}\tnote{b} & $9.32$ & $2.95$ \\
HECG (present)\tnote{*} & $8.7(2)$ & $3.2(1)$ \\
Oda \emph{et al.}~\cite{oda2001}\tnote{c} & $8.26$ & $3.02$ \\
Platzmann and Mills~\cite{platzman1994} & $\approx 5.7$ & $\approx 1.9$ \\
QMC~\cite{shumway2001}\tnote{d} & $9.148(42)$ & $3.024(58)$ \\
QMC~\cite{shumway2005}\tnote{e} & $\gtrsim9.148(42)$ & $2.900(34)$ \\
Superseded SVM~\cite{ivanov2001,ivanov2002}\tnote{f}& $8.443$ & $2.998$ 
\end{tabular}
\end{ruledtabular}
\begin{tablenotes}
\item[a] For basis set Ps$(1s)$Ps$(1s,2s,2p)$.
\item[b] For basis set Ps$(1s,2s,2\bar{p},3\bar{d})$
Ps$(1s,2s,2\bar{p},3\bar{d})$; 
bar denotes pseudostate.
\item[c] Model: long-range van der Waals potential with short-range hard core, 
constrained to fit Ps$_2$ binding energy.
\item[d] Polynomial fit to phase shift.
\item[e] Same data as~\cite{shumway2001}, but fit to effective range theory.
Singlet value unknown, but deduced to be slightly larger 
than in~\cite{shumway2001}.
\item[f] Error bars beyond digits shown.
\item[*] See text for the description of the error estimate.
\end{tablenotes}
\end{threeparttable}
\end{table}
This systematic error at low scattering energy arises from
a number of sources.
As already mentioned,
the oscillations in $\tan\delta$ obscure the underlying
linear behavior if the $R$-matrix is not propagated 
to sufficiently large distance.
The more probable cause is that
at low scattering energy
the scattering data is very sensitive to small changes 
in the realistic potentials, 
that is, the adiabatic potentials $U_{\nu}(R)$ 
with diagonal $Q_{\nu\nu}(R)$ correction.
In practice,
where possible,
an inverse power-law fit is performed on the large-$R$ tails 
of the realistic potentials as this matches the expected behavior
at large distance.
However,
in some cases a nonadiabatic coupling occurs in the region where
the inverse power-law asymptotics would be expected,
obscuring this behavior.
Thus,
instead of a power-law fit,
we find a Lorentzian-like tail to be a more appropriate large $R$ fit
to the realistic potential.
Nevertheless,
since the realistic potentials are a variational upper bound to the true
potentials,
the scattering data at low scattering energy is systematically higher
than expected.
A fit of the zero-energy $s$-wave scattering lengths using 
the scattering data at higher energy would provides an estimate much
closer to accepted values,
but is not provided here.

We estimate the error by fitting the $N$-channel scattering data 
to quadratic and linear polynomials,
extrapolating to zero scattering energy.
An example is shown in Fig.~\ref{fig_elasticscattering}
for the inifite channel approximation (crosses).
This provides error estimates for each fixed-channel calculation.
The difference between the 4-channel calculation and infinite-channel approximation
are used to then estimate an overall error.

The scattering lengths reported in Table~II 
are not the experimentally relevant scattering lengths.
Those reported are for the singlet-singlet
or triplet-triplet symmetries of the identical electrons and positons.
For experiment,
it is the spin of the Ps atoms that is relevant.
It is a straightforward calculation to switch between the two
coupling schemes; see, e.g. Refs.~\cite{shumway2005} or~\cite{ivanov2002}.

\section{Charge redistribution}
\label{sec_charge}
This section describes the charge redistribution,
that is,
the probability of transferring from a dimer-dimer channel
to the ionic channel
that occurs in only the systems with SS symmetry.
The probabilities are given by the squared absolute value of the 
$S$-matrix elements.
The $S$-matrix is calculated after propagating the $R$-matrix to a large distance.
Similar to the approach shown in Fig.~\ref{fig_phaseshift},
each element of the $S$-matrix is extrapolated to matching to the asymptotic
solutions at $R\to\infty$.

In the following,
for positive charge conjugation symmetry,
only the lowest four channels are included in the calculations
[the solid, dashed, dash-dotted, and dash-dash-dotted lines of Fig.~\ref{fig_CplusSSTT_scaled}(a)].
The dimer-dimer channels fragment 
into one ground state $s$-wave Ps and one excited state $s$-wave Ps.
No $d$-wave [dotted and upper solid lines of Fig.~\ref{fig_CplusSSTT_scaled}(a)]
or the highest $s$-wave [dash-dot-dotted line of Fig.~\ref{fig_CplusSSTT_scaled}(a)]
fragmentation channels are included.
Beyond the crossing with the third dimer-dimer channel,
the ionic channel is made continuous up to the ionic threshold of $-0.262E_H$,
where all crossings with the neglected dimer-dimer channels are assumed to be
fully diabatic.
For the negative charge conjugation symmetry,
only the lowest three channels are included in the calculations
[the solid, dashed, and dash-dotted lines of Fig.~\ref{fig_Cminus_scaled}(a)].
The asymptotically dimer-dimer channels fragment 
into one ground state $s$-wave Ps and one excited state $p$-wave Ps.
No $f$-wave [not shown in Fig.~\ref{fig_Cminus_scaled}(a)]
or the highest $p$-wave [dash-dash-dotted line 
of Fig.~\ref{fig_Cminus_scaled}(a)]
fragmentation channels are included.
Beyond the crossing with the second dimer-dimer channel,
the ionic channel is made continuous up to the ionic threshold of $-0.262E_H$,
where all crossings with the neglected dimer-dimer channels are assumed to be
fully diabatic.

\begin{figure}
\vspace*{+1.5cm}
\centering
\includegraphics[angle=0,width=70mm]{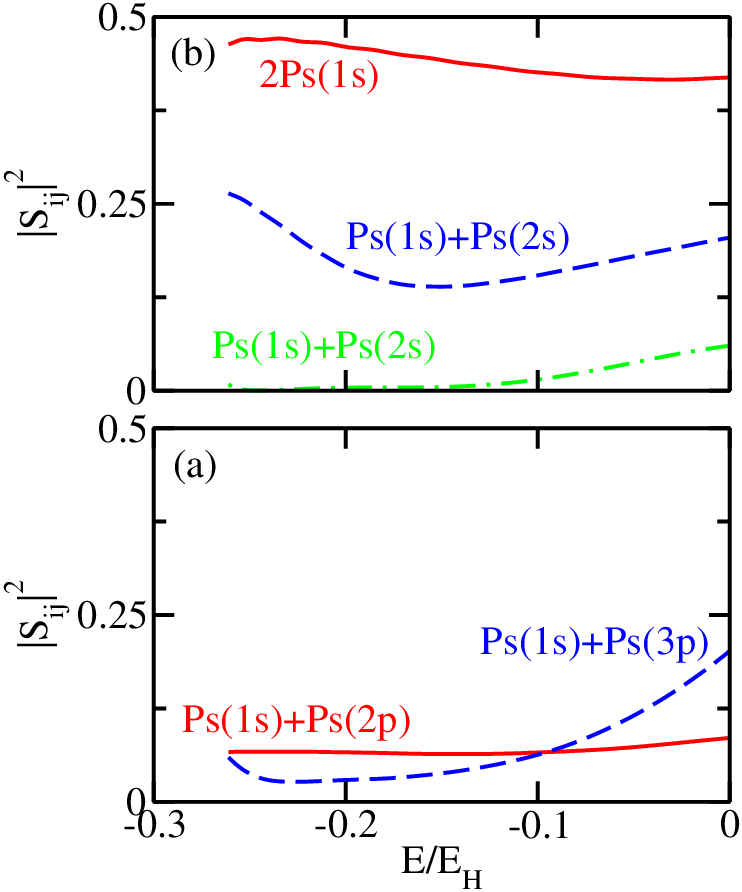}
\vspace*{0.1cm}
\caption{(Color online)
Charge redistribution probabilities as a function of scattering energy for SS and 
(a) positive or (b) negative charge conjugation symmetry.
All curves are for dimer-dimer to ionic transitions
and are labeled by the dimer-dimer threshold.
The ionic threshold is $-0.262E_H$.
}\label{fig_charge}
\end{figure}

Figure~\ref{fig_charge}
shows some of the squared absolute values of the off-diagonal $S$-matrix elements
for the SS system with (a) positive and (b) negative charge conjugation symmetry.
In panel (a),
solid ($|S_{14}|^2$), dashed ($|S_{24}|^2$), and dash-dotted ($|S_{34}|^2$) lines
show the transition probabilities to transition from
the three lowest dimer-dimer channels to the ionic channel.
The charge transfer is most efficient from the ground dimer-dimer channel
and is less efficient as the excited Ps increases in principle quantum number.
This can be understood since the avoided crossing is largest between the ground
and first excited states.
Thus, 
any flux coming in on the ground state is efficiently transferred to the higher channels
as opposed to simply exciting one of the Ps atoms,
whereas flux coming in on the first excited state is 
more efficiently given to the ground state
rather than transferring to the ionic channel.
The crossing of the third dimer-dimer channel is almost fully diabatic,
hence there is no efficient charge transfer from this state to
the state of the Ps ion and a free charge.
This trend of the crossings becoming more diabatic as the energy increases
supports our approach of neglecting the more excited dimer-dimer channels.

In Fig.~\ref{fig_charge}(b),
solid ($|S_{13}|^2$) and dashed ($|S_{23}|^2$) lines
show the transition probabilities to transition from
the two lowest dimer-dimer channels to the ionic channel.
Overall, the charge transfer is not as efficient in comparison to the
case of positive charge conjugation symmetry.
This can be understood since there is only one wide avoided crossing, 
but it is not as wide as in the case of positive charge conjugation symmetry.
The second curve crossing is already mostly diabatic,
suppressing much of the probability to transfer to the ionic channel.

\section{Conclusion and Outlook}
\label{sec_conclusion}
This paper calculates the lowest adiabatic potential curves 
as a function of the hyperradius $R$ for
the $2e^+2e^-$ system for zero orbital angular momentum,
positive parity,
and different charge conjugation symmetries.
Using these hyperradial channels,
low-energy elastic scattering properties are determined
by propagating the $R$-matrix.
The long-range couplings from the Coulomb interactions are overcome
by matching to the asymptotic hyperradial functions at
many different points.
The observed behavior in the tangent of the phase shift
and the calculated $S$-matrix elements,
as a function of inverse matching point,
is linear with damping oscillations.
The resulting $s$-wave scattering lengths are larger
than other values in the literature,
but nevertheless
converge well as a function of the number of included channels
and provide reasonable estimates of the scattering properties.
The transition probabilities are expected to be good estimates,
but could be improved by including more channels.

The ability to treat the ionic and dimer-dimer fragmentation channels
on an equal footing is one of the strengths of the adiabatic
hyperspherical method.
Future studies will extend this system to include inelastic scattering
properties and different orbital angular momentum states.
Moreover,
the masses and charges of the particles are tunable parameters.
It would be interesting to study e.g. the change in the potential
curves and transition amplitudes as a function of the
mass of the positive charge
transitioning from the Ps$_2$ to the H$_2$ system.
Explicitly including hydrogenic wave functions in the basis
could also help the convergence issues at large hyperradius.
These topics will be the focus of future studies.

\section{Acknowledgements}
Support by the National Science Foundation through Grant No. PHY-1306905
and by the US Deptartment of Energy, Office of Science through Grant No. 
DE-SC0010545 is gratefully acknowledged.

\end{document}